\documentclass[twocolumn]{emulateapj}

\usepackage{graphicx}
\usepackage{amsmath, amsthm, amssymb}
\usepackage{natbib}
\usepackage{epsfig}
\usepackage{subfigure}
\usepackage[colorlinks = true, linkcolor=blue, citecolor=blue, urlcolor=blue]{hyperref}

\usepackage{aas_macros}
\usepackage{natbib}
\usepackage{color}
\defcitealias{2008ApJ...675.1125B}{B08}
\defcitealias{2010A&A...513A..37H}{H10}

\shorttitle{Probing the Galaxy Cluster Merger in A115}
\shortauthors{Hallman et al.}

\begin{document}


\title{Probing the Curious Case of a Galaxy Cluster Merger in Abell 115 with High Fidelity Chandra X-ray Temperature and Radio Maps}


\author{Eric J. Hallman$^{\dagger}$, Brian Alden, David Rapetti\altaffilmark{a}, Abhirup Datta\altaffilmark{b}, Jack O.\ Burns}
\affil{Center for Astrophysics \& Space Astronomy, Department of Astrophysical \& Planetary Sciences,
	389 UCB, University of Colorado, Boulder, CO 80309, USA}
\email{$^{\dagger}$eric.hallman@colorado.edu}

\altaffiltext{a}{Also at NASA Ames Research Center, Moffett Field, CA 94035, USA}
\altaffiltext{b}{Also at Indian Institute of Technology, Indore, India}


\begin{abstract}
We present results from an X-ray and radio study of the merging galaxy cluster Abell 115. We use the full set of 5 Chandra observations taken of A115 to date (360 ks total integration) to construct high-fidelity temperature and surface brightness maps. We also examine radio data from the Very Large Array at 1.5 GHz and the Giant Metrewave Radio Telescope at 0.6 GHz. We propose that the high X-ray spectral temperature between the subclusters results from the interaction of the bow shocks driven into the intracluster medium by the motion of the subclusters relative to one another. We have identified morphologically similar scenarios in Enzo numerical N-body/hydrodynamic simulations of galaxy clusters in a cosmological context. In addition, the giant radio relic feature in A115, with an arc-like structure and a relatively flat spectral index, is likely consistent with other shock-associated giant radio relics seen in other massive galaxy clusters. We suggest a dynamical scenario that is consistent with the structure of the X-ray gas, the hot region between the clusters, and the radio relic feature. 
\end{abstract}



\section{Introduction}

In a cold dark matter dominated universe, structures form hierarchically, leading to mergers of smaller gravitationally bound objects into bigger ones. When the most massive structures -- galaxy clusters -- merge, they result in the most energetic events in the universe since the Big Bang. Observational evidence of shocked gas in merging galaxy clusters is now relatively common \citep{bullet, Markevitch:07, canning, emery}. In particular, X-ray surface brightness and temperature maps show features that strongly suggest they result from shocks driven by the supersonic motion of merging subclusters. A shock in the intracluster medium (ICM) will result in both a density and temperature discontinuity in the gas, which creates an X-ray excess and a spectral temperature jump. Shocks can also manifest observationally in the form of radio ``relics", arc-like structures in the outskirts.  They are believed to be the result of shocks induced by mergers, compression of radio lobes, or the remnants of radio galaxies \citep[for recent reviews, see][]{bruggenReview,bruJones}. Except in a few cases \citep[e.g.,][]{finoguenov,Datta:14} shocks in the ICM associated with radio relics are not detectable by their X-ray emission, due to their location far from the cluster center (r $>$ 1 Mpc) where the X-ray surface brightness is too low to reliably detect an enhancement \citep{hoeft07}. Prior work \citep{Botteon:16} suggests that Abell 115 may also be an example where there is an X-ray shock at the location of the relic.

A reasonable expectation is that all merging galaxy clusters will contain shocks \citep{ryu03,skillman08,vazza11}, some of which may be observable through X-ray observations, modulo orientation effects that may smear out the contrast across the steep, narrow pressure discontinuity, and the X-ray surface brightness in the local gas.  

\subsection{Abell 115}
Abell 115 is a well-known massive (total virial mass $M_v \approx 3\times10^{15} M_\odot$) galaxy cluster at redshift $z = 0.192$, with double peaked structure in the X-ray surface brightness \citep{Forman:81,Shibata:99,Gutierrez:05}.  Optical studies \citep{Barrena:07} indicate two distinct, redshift-separable components in the local galaxy population, roughly coincident with the two X-ray peaks. Both subclusters have disturbed morphology in the X-ray, and host cool cores, where some of the cool gas appears to be in the process of being stripped from the cluster core. X-ray and optical observational data, therefore, are consistent with the interpretation that the two subclusters are in the process of a merger. 
In addition, the northern subcluster hosts a 3C radio source at its center, 3C28 \citep{forman10}. As has been noted by prior studies, to the northeast of 3C28, there exists extended, diffuse radio emission in an arc-like structure that seems consistent with the appearance of other so-called cluster radio ``relics" \citep{govoni:01,Botteon:16}. While some of the radio emission appears to come directly from discrete radio galaxies, there is significant radio structure stretching between these individual sources. One possible interpretation of the extended radio structure is that this emission results from particle acceleration at a shock, or shocks, driven into the ICM by the motion of the subclusters relative to one another. Arc-like relics exist in other massive galaxy clusters \citep[e.g.,][]{rottgering,bonafede,giovannini,sausage}, and in many cases, a convincing argument has been made that the emission results from synchrotron radiation of a Fermi-accelerated particle population in the ambient intracluster magnetic field.  Indeed, prior work on A115 has made a similar argument \citep{Botteon:16}, citing evidence in the X-ray data for a shock feature coincident with the location of the extended radio structure to the northeast of 3C28.

Until now, a detailed description of the dynamics of A115 that is consistent with all of the multi-wavelength observations of A115 has not been offered. In this work we deduce the dynamics of A115 using evidence from the X-ray, radio and optical observations, in addition to comparison with numerical simulations. In Section \ref{sec:Chandra}, we discuss the data reduction we used for the Chandra X-ray observations of A115. In Section \ref{sec:RadioData}, we describe the data reduction of the VLA and GMRT radio observations. In Section \ref{sec:StrucAndDyn}, we give a description of the X-ray temperature structure and the likely dynamics of the cluster. In Section \ref{sec:RadioRelic}, we combine our evidence from both X-ray and radio data to interpret the location and morphology of the radio relic. In Section \ref{sec:Discussion}, we summarize the results and suggest future work.
\section{Chandra X-ray Data Reduction}\label{sec:Chandra}
We used multiple \textit{Chandra} observations in our
analysis. The \textit{Chandra} observations (IDs 3233, 13458, 13459, 15578, 15581) were obtained from the Chandra Data Archive. The observations for 3233 were taken in 2002, while the later 4 were taken in November of 2012. Exposure times were $\sim$50 ks, $\sim$115 ks, $\sim$100 ks, $\sim$65 ks and $\sim$30 ks respectively, for a combined total of roughly 360 ks across the 5 observations.  All were observed in VFAINT mode. 

\begin{figure*}[htbp] 
   \centering
   \includegraphics[width=0.7\textwidth]{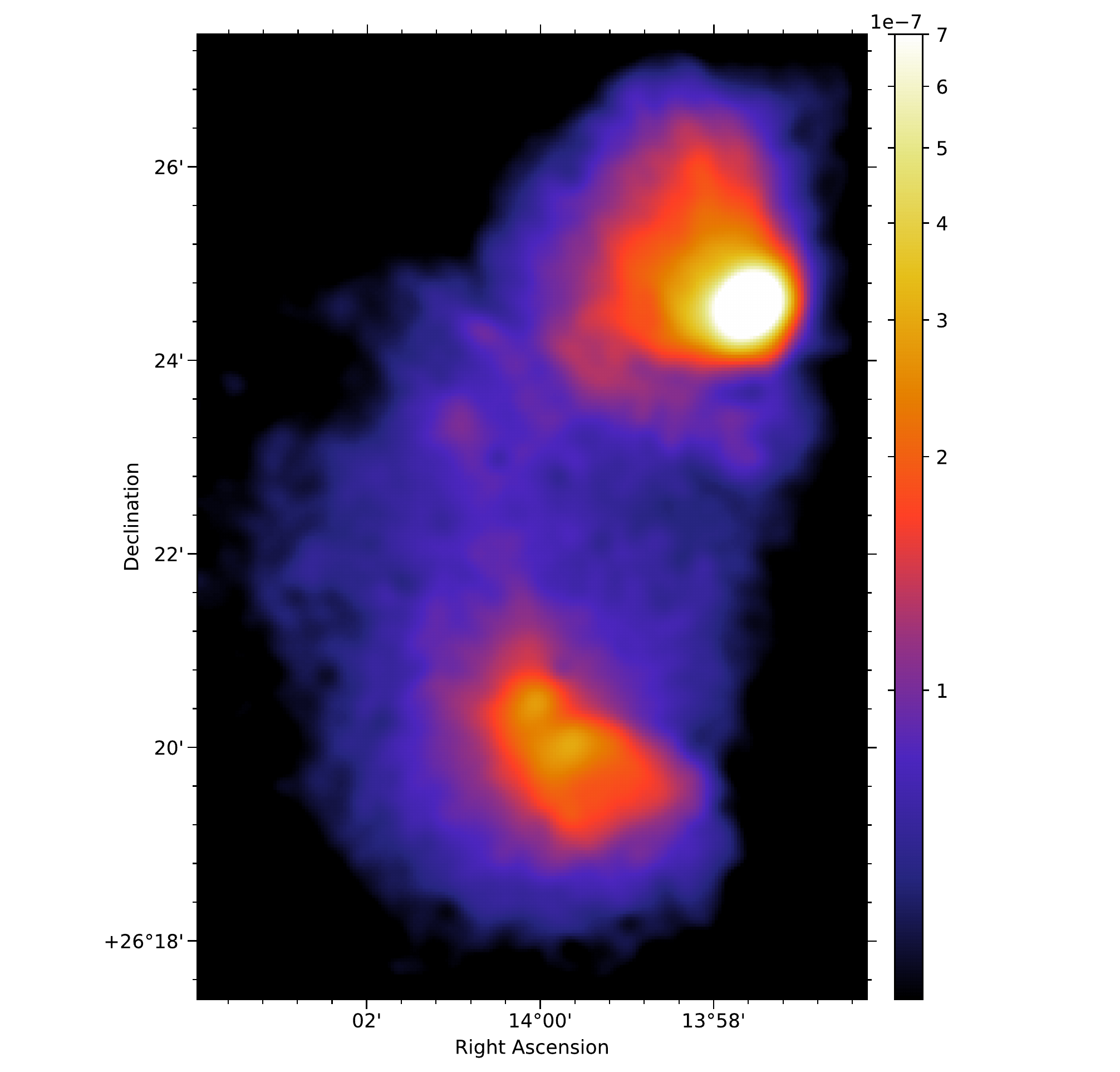}
   \caption{Combined, background-subtracted, smoothed (1.5" FWHM), 0.5-8.0 keV X-ray image constructed from all 5 \textit{Chandra} observations used in this study. Point sources have been excised. Units are counts/cm$^2$/s.}
   \label{fig:fluxFig}
\end{figure*}

\subsection{The X-ray ``Pypeline" for Data Reduction and Temperature Maps}

The X-ray data reduction is described in detail in \citet{Schenck:14} and \citet{Datta:14}. The data reduction described in these studies has been aggregated into a data pipeline, which is designed to take \textit{Chandra} observations and generate high resolution adaptive circular binned (ACB) temperature maps. Once \textit{Chandra} observation IDs are given as input, the pipeline automatically downloads the data using \textit{CIAO}\footnote{http://cxc.harvard.edu/ciao/} and merges the multiple observations into a single image. Currently, the end user needs to provide a SAO DS9\footnote{http://ds9.si.edu/} region file containing the point sources for exclusion. Once the sources file is given to the pipeline, it removes the sources from the images. The pipeline then generates light curves for each observation and removes flares. The user has the ability to inspect and customize this process to ensure accuracy. Response files are then generated for each observation using \textit{specextract}. To create the ACB temperature map, $~\mathcal{O}(10^4)$ spectral fits are generated. This part of the process can be done in parallel on a supercomputer to drastically reduce the time required to complete the map. 

The ACB temperature maps were produced using a method adapted from \citet{Randall:08, Randall:10}. In these two papers, spectra were extracted from circular regions which were just large enough to reach some threshold of counts. The fitted temperature of each region was assigned to the pixel at the center of the circle. The circles are allowed to overlap,
so some pixels will share counts with other pixels and the fitted temperatures
will not be independent from one another.

The pipeline itself is written in Python and available for the community at large as an open source project on GitHub\footnote{https://github.com/bcalden/xray-tmap-pypeline}. Future desired functionality, includes native python multi-threading support, an automatic source finder, and graphical user interface improvements. The pipeline is currently in beta.

\subsection{X-ray Surface Brightness and Temperature Image Generation}
The \textit{Chandra} data were calibrated using \textit{CIAO} 4.7 and
\textit{CALDB} 4.7, the most up-to-date versions at the time of analysis. Bad
pixels and cosmic rays were removed using
\textit{acis}\_\textit{remove}\_\textit{hotpix} and charge transfer inefficiency (CTI) corrections were made
using \textit{acis}\_\textit{process}\_\textit{events}. Intervals of background
flaring were excluded using light curves in the full band and the 9-12 keV band.
The light curves were binned at 259 seconds per bin, the binning used for the
blank-sky backgrounds. Count rates greater than 3-$\sigma$ from the mean were
removed using \textit{deflare}. We visually inspected the light curves to
ensure flares were effectively removed. We used the blank-sky backgrounds in
\textit{CALDB} 4.7. The backgrounds were reprojected and processed to match the
observations.  Figure \ref{fig:fluxFig} shows the combined, background-subtracted, point source excised 0.5-8.0 keV surface brightness image.  

The ACB temperature map is created by fitting a thermal plasma model to the multiple extracted X-ray spectra.
For \textit{Chandra}, the source and background spectra were extracted using
\textit{dmextract} and the weighted response was extracted using
\textit{specextract}. We rescaled the background spectra using the ratio of the
high-energy counts (9.5-12 keV) in the source and the background. We expect the
counts to be predominantly from the background at these high energies.
\begin{figure*}[htbp] 
   \centering
   \includegraphics[width=0.49\textwidth]{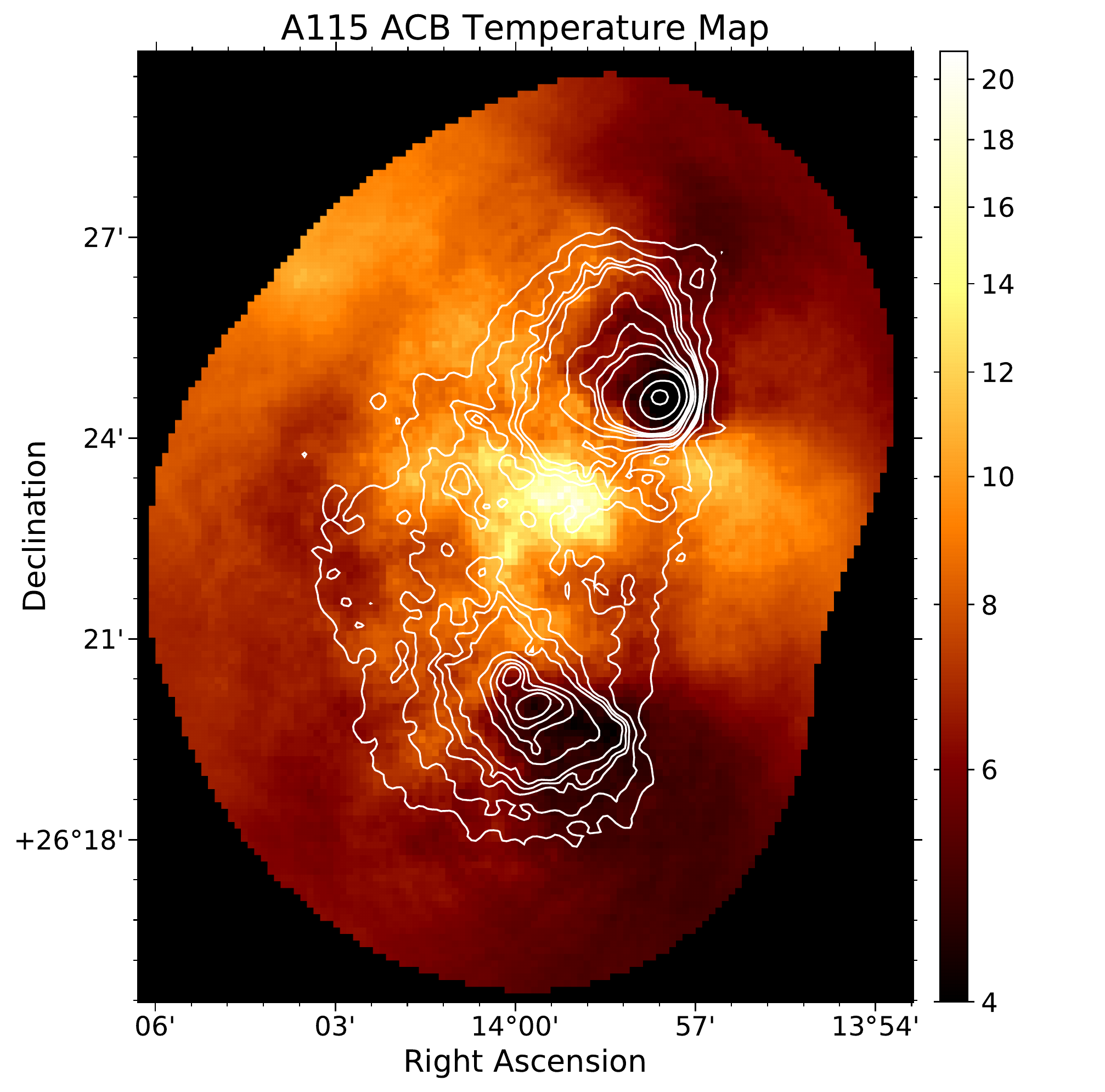}
   \includegraphics[width=0.49\textwidth]{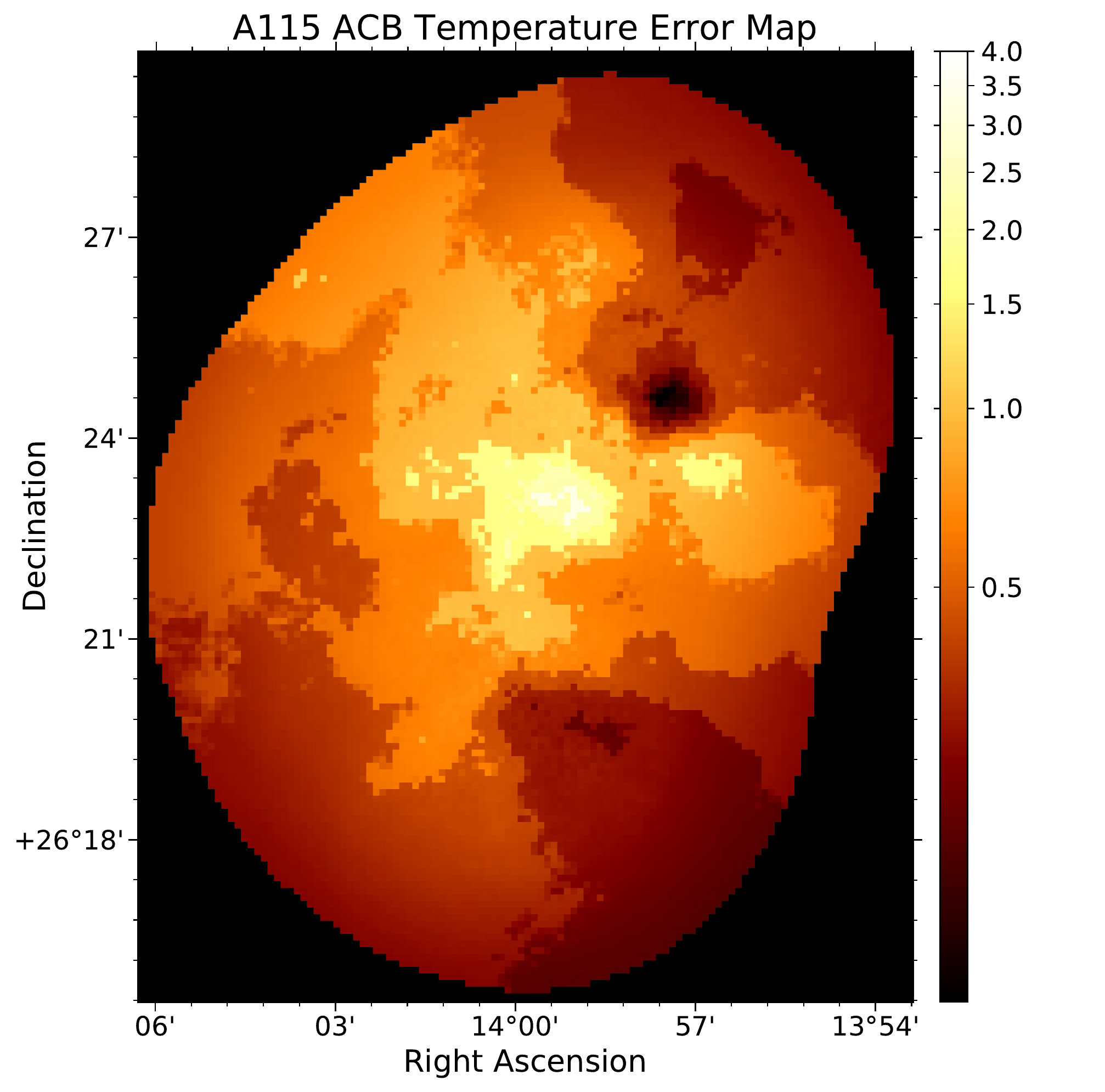}
   \caption{The left image is an X-ray temperature map constructed with the adaptive circular binning (ACB) technique described in the text, overlaid with 11 X-ray surface brightness contours at 1, 1.4, 1.8, 2, 2.2, 3.2, 4, 4.6, 5.2, 8, and 20 percent of peak emission ($\sim 5 \times 10^{-6}$ counts/cm$^2$/s).  The right image is the  error (1 $\sigma$) associated with the temperature map. These images are used as a guide to help better understand the dynamics of the cluster. Both images are measured in keV.}
\label{fig:acbMap}
\end{figure*}

The spectra were then fit using an \textit{APEC} thermal plasma model in \textit{XSPEC}\footnote{https://heasarc.gsfc.nasa.gov/xanadu/xspec/}. We included photoelectric absorption, but the Galactic hydrogen column density was not fit. Instead, it was frozen at a value of 5.2$\times$ $10^{20}$ cm$^{-2}$ \citep{starkHI}. The resulting temperature map, with X-ray surface brightness contours, is shown in Figure \ref{fig:acbMap}.

\section{Radio Data}\label{sec:RadioData}
So far, the published radio map for A115 has been by \citet{Botteon:16} for VLA C+D array data (also produced here as red contours in Figure~\ref{fig:SBandRadio}).
Here, we have used archival multi-frequency radio observations of Abell 115 (see Table \ref{obs_journal}) in order to investigate on the nature of the diffuse radio emission in the cluster.

\subsection{GMRT 610 MHz}

GMRT 610 MHz archival data on A115 were analysed in \textit{CASA} (see Table~\ref{obs_journal}). The data were downloaded in FITS format. We converted the FITS file into \textit{CASA} MS format through the \textit{CASA} task \textit{importgmrt}. First, non-functional antennas were flagged based on the observing log. Then we used \textit{AOFlagger}\footnote{http://aoflagger.sourceforge.net} \citep{Offringa:12} for radio frequency interference (RFI) flagging. \textit{AOFlagger} is a  framework that implements several methods to deal with RFI. About 30\% of the data were flagged in \textit{AOFlagger}. Then in the output MS, further manual flagging was done. 12 out of 30 antennas and 73 out of 256 channels were flagged including the frequency band edges. After flagging some outliers and clipping some bad data, the rest were calibrated with the standard calibrator 3C48 (used as both flux and phase calibrator). Then, we separated the calibrated target data with the task \textit{split}. 

\begin{table}
\caption{Summary of the Archival Radio Observations}
\label{obs_journal}
\begin{tabular}{|c|c|c|c|c|c|}
\hline
Array & $\nu$ & 	BW & Project &	Date of & PI\\
~ & (MHz) & (MHz) & Code & Obs. & ~\\
\hline

GMRT &	610  &	32 & 20-004 &	30.6.11  & A. Bonafede\\
VLA-D &	1420	& 50	 & AF0349 & 	26.3.99 & L. Feretti\\
VLA-B &	1420 & 64	& 15A-270	& 15.2.15 & R. van Weeren\\

\hline\hline
\end{tabular}
\end{table}

We tried to recover diffuse emission from the GMRT 610 MHz data at the region between 3C28 and the head tail source as seen in Figure~\ref{fig:SBandRadio}. However, at the fullest resolution of the GMRT 610 MHz data we could hardly recover any diffuse emission. Hence, we have selected only the $0$ to $\sim$15 $k\lambda$ u-v range of GMRT data at 610 MHz for imaging, which recovered some amount of diffuse emission in the bridge region as now seen in Figure~\ref{fig:VLA-GMRT}. The imaging was performed by the \textit{CASA} task \textit{clean} choosing a cell size of 9$\arcsec$. Briggs weighting was again used with robust parameter -1. Wide-field imaging was done in \textit{CASA} using w-projection algorithm with 512 w-projection planes. The image was restored with a beam 45$\arcsec$ $\times$ 45$\arcsec$. The RMS noise near the center of the field is $\approx$ 1.4 mJy/beam.

\subsection{VLA L-band D-configuration} 
\begin{figure*}[htbp] 
   \centering
   \includegraphics[width=0.7\textwidth]{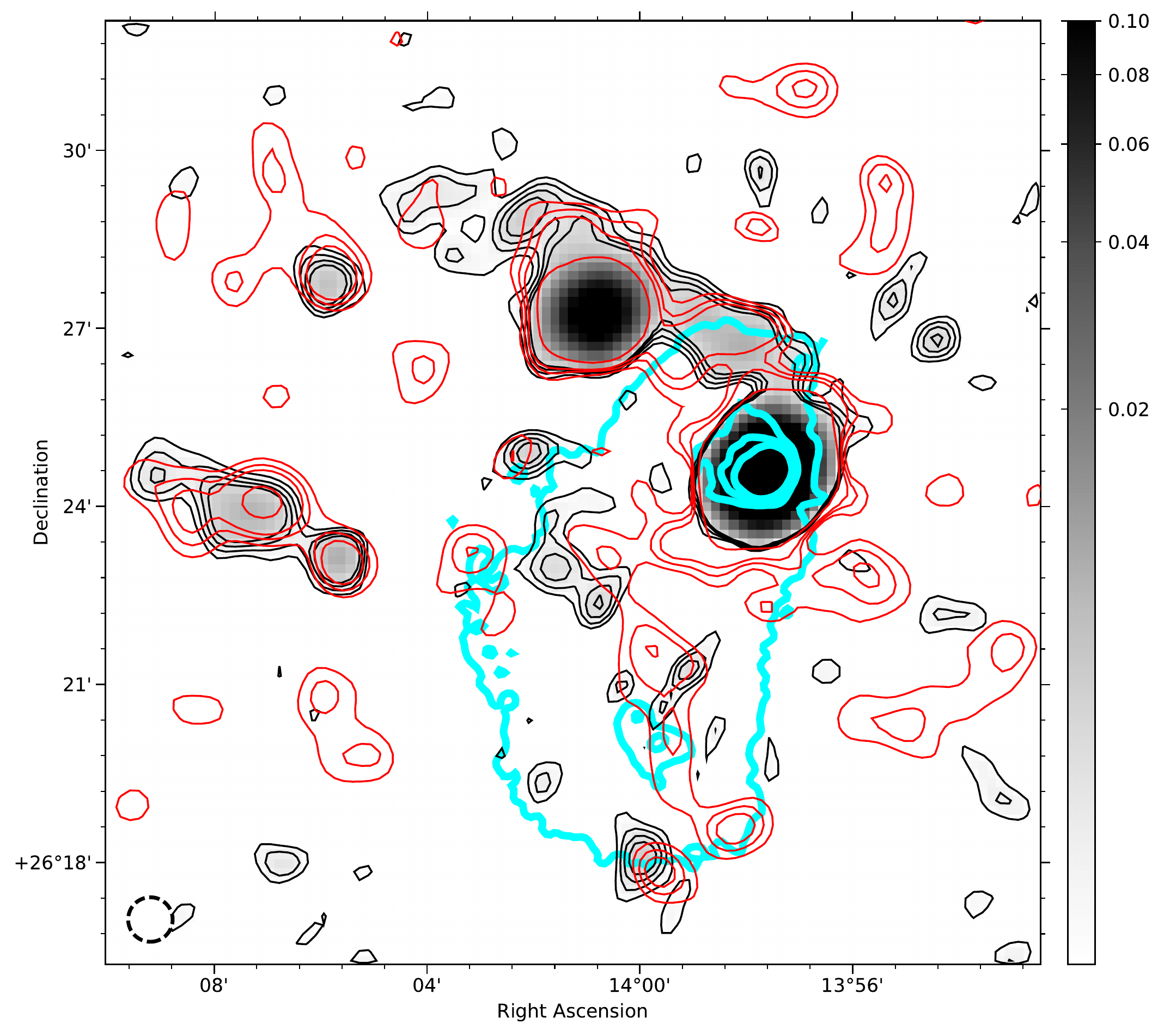}
   \caption{The color image shows the VLA-D array image with restoring beam size of 45$\arcsec$ (depicted by the dashed black circle in the bottom left). The off-source RMS noise is $\approx$500 $\mu$Jy/beam near the center of the field. The black contours are also from VLA-D array observations and represent 0.21, 0.31, 0.41, and 0.52 percent of peak emission. The red contours are for GMRT 610 MHz observations and are at 0.18, 0.29, 0.40, and 0.88 percent of peak emission. The thick cyan contours depict 1, 4, 6, and 8 percent of the peak emission in X-ray surface brightness. It should be noted that the GMRT results are with the same restoring beam sizes of 45 arcsec FWHM. The off-source RMS noise is about 1.4 mJy/beam near the center of the field. The colorbar is in Janskys/beam.}
   \label{fig:VLA-GMRT}
\end{figure*}

The A115 VLA L-band D configuration archival data were analyzed in \textit{CASA}\footnote{https://casa.nrao.edu/}. The data were converted into measurement set (MS) format via task \textit{importvla}. Then in the output MS file we have applied manual flagging. 7 out of 27 antennas were flagged. Then the calibration was completed using standard flux calibrator 3C48 and phase calibrator 0119+321. The calibrated target field data was separated from the multi-source dataset by the task \textit{split} choosing only RR and LL correlations. Imaging was performed by the \textit{CASA} task \textit{clean} with the imaging mode `channel' as this archival data contains single channel per IF. The VLA D-configuration synthesized beam size in L-band is 45$\arcsec$, so we chose the cell size to be 9$\arcsec$. Briggs weighting was used with robust parameter -1. The image was restored with a Gaussian beam with a 45$\arcsec$ FWHM. The RMS noise is $\approx$500 $\mu$Jy/beam near the center of the field. The results from VLA-D array L-band are shown in Figures~\ref{fig:VLA-GMRT} and \ref{fig:SBandRadio}.  In order to derive spectral index between any two images at two different radio frequencies, the images from both radio frequencies need to be of same angular resolution. Since the restoring beam of GMRT 610 MHz image is 45$\arcsec$ $\times$ 45$\arcsec$, we needed to match that resolution in VLA-D array L band as well.

\subsection{VLA L-band B-configuration}
The Abell 115 L band B array VLA data were first run through the standard \textit{CASA} calibration and editing processes. The editing tasks like \textit{flagdata} and \textit{rflag} were used to flag the bad data. 3 out of 27 antennas along with some channels were flagged. The calibration was done using tasks \textit{setjy} and \textit{gaincal}. The gain solutions were checked using \textit{plotcal}. The calibrated target data were separated from the multi-source data set with the task \textit{split}. The imaging was done by the \textit{CASA} task \textit{clean} choosing a cell size of 1 arcsec. Briggs weighting was used with robust parameter -1. Wide-field imaging was done in \textit{CASA} using the w-projection algorithm with 128 w-projection planes. The image was restored with restoring beam 3.2$\arcsec \times 2.8\arcsec$ with a beam p.a. of 78$^o$. The RMS noise is $\approx$45 $\mu$Jy/beam near the center of the image. The results from VLA-B array L-band are shown in Figure~\ref{fig:SBandRadio}. It is evident from this Figure that the resolution of the VLA-B array helps us to resolve the structure in the two sources: the well-known 3C28 (in the west) and the head-tail radio source (J0056+2627) to the northeast.

\subsection{Radio Spectral Index and Mach Number}
It is evident from the Figure~\ref{fig:VLA-GMRT} that the diffuse emission is best captured with the VLA-D array and GMRT 610 MHz analysis with broader restoring beam. We then calculate the spectral index in the for the bridge of radio emission between 3C28 and the head-tail radio source (J0056+2627) to the northeast, we get the average spectral index to be $-1.1 \pm 0.2 $. With a 45$\arcsec$ restoring beam, the bridge is unresolved in the transverse direction. However, the quality of the archival radio data prevents us from going any further with this analysis and creating a spectral index map of the region. In order to do so, we need new radio observations of this field at L and P bands with the VLA. The upgraded capabilities of the VLA will allow us to get a spectral index map of this 'bridge'. 

If we are viewing a shock front oriented edge-on, the radio spectral index ($\alpha$) should be sensitive to the prompt emission from the shock front \citep{skillman13}, given by $\alpha=\alpha_{prompt} = (1-s)/2$, where $s$ is the spectral index of the accelerated electrons given by $n_e(E) \propto E^{-s}$ \citep{hoeft07}. The theory of diffusive shock acceleration (DSA) for planar shocks, at the linear test-particle regime, predicts that this radio spectral index is related to the shock Mach number by:
\begin{equation}\label{eq:radmach}
M^2=\frac{2\alpha-3}{2\alpha+1}
\end{equation}
where $\alpha$ is the radio spectral index ($S_\nu \propto \nu^\alpha$) \citep{blandford87,hoeft07, ogrean13}.  
Given the orientation of the features, we estimate a Mach number for the edge-on case. For prompt emission case and a spectral index within the relic region of A115 ($\alpha = -1.1$), the resulting Mach number is $2.1$. This is consistent with the estimates of the shock Mach number at the relic position computed by \citet{Botteon:16} using the X-ray data.

\section{X-ray Temperature Structure and Dynamics}\label{sec:StrucAndDyn}
The X-ray surface brightness morphology, coupled with the high-fidelity temperature map, strongly suggest a likely dynamical scenario, at least as projected on the sky.  The addition of the optical redshifts of the member galaxies indicates the relative line of sight motion as well. \citet{Barrena:07} analyzed optical redshift and photometric data for 115 galaxies, all of which were members of A115-N and A115-S. Their analysis strongly indicates two separate distributions of galaxies, A115-N with a velocity dispersion of $\sigma_v \approx 1000$ km s$^{-1}$, and A115-S with $\sigma_v \approx 800$ km s$^{-1}$. The analysis suggests the subclusters are moving toward one another along the line of sight with relative velocity $V_r \approx 1600$ km s$^{-1}$. 

The defining features of the X-ray temperature map, shown in Figure \ref{fig:acbMap}, are the two cold, bullet-like structures in the north and south, with temperatures in the cold gas of $3.5 \leq T_X \leq 5$ keV, and the hot, amorphous region between them, with X-ray temperatures as fit in the ACB map ranging up to 15-20 keV. The overall dynamical picture seems to be that the two subclusters are moving both toward one another along the line of sight (as suggested by the optical data), and moving past each other in the plane of the sky (as suggested by the X-ray data). Given the elongated X-ray appearance, the northern subcluster appears to be moving to the west and slightly south, while the southern is moving to the north and east. We explored this interpretation by extracting spectra from the data. 

\begin{figure*}[htbp] 
   \centering
   \includegraphics[width=0.7\textwidth]{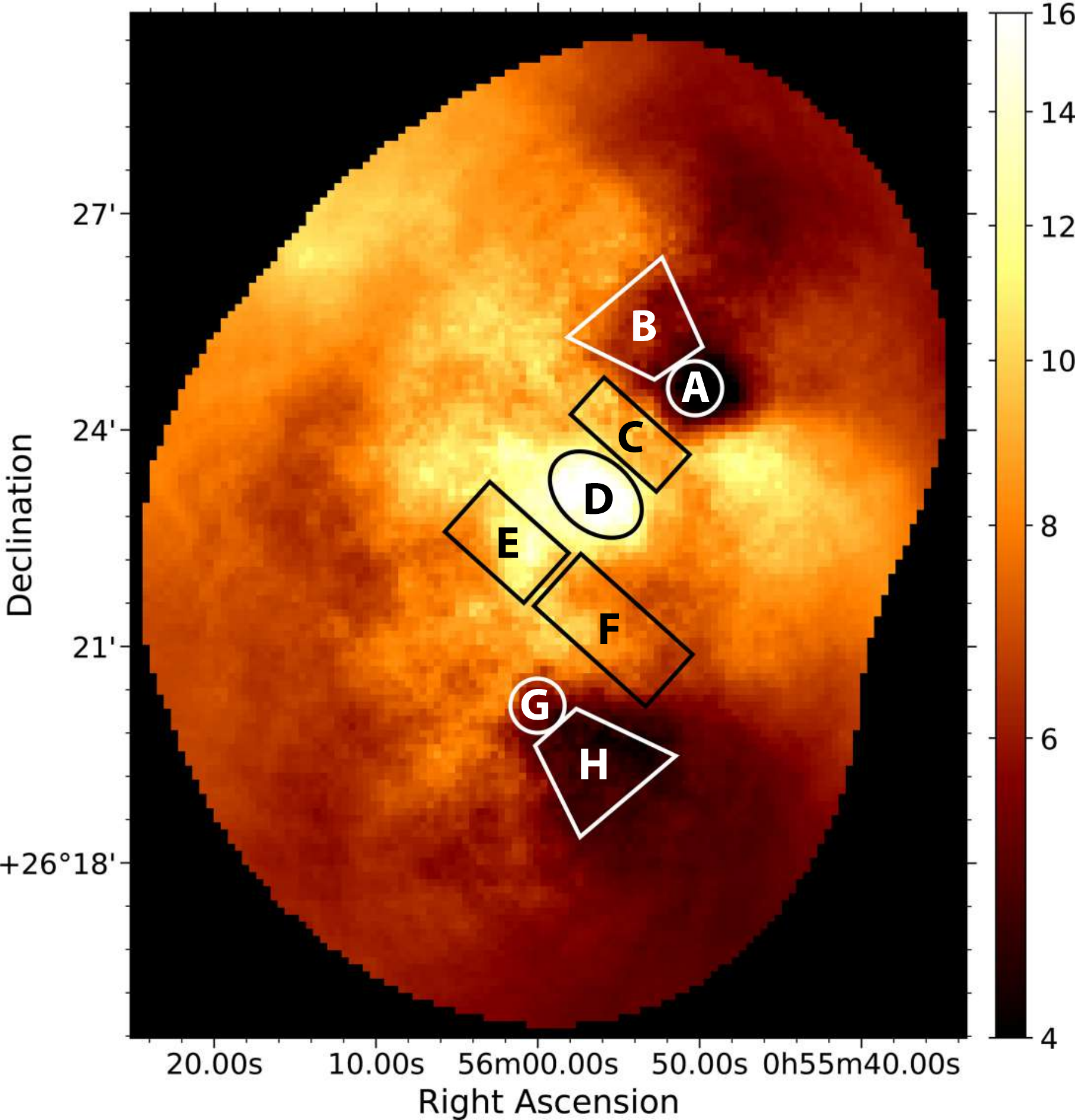}
   \caption{Image of the ACB temperature map, created as described in the text. Overlaid on the image are the regions from which spectra have been extracted and fit, using a simultaneous fit to all observations. Color scale indicates spectral temperature in units of keV. See Table \ref{XrayFits} for spectral temperature fits of these regions.}
   \label{fig:regions}
\end{figure*}

In Figure \ref{fig:regions}, we show a set of regions, guided by features in the ACB map, that we have extracted spectra from (using \textit{specextract}), and fit with \textit{XSPEC}. The regions were chosen to isolate areas of interest in the ACB temperature map. Regions A and G contain the cold, X-ray bright cores of A115-N and A115-S respectively.  Regions B and H were chosen to cover what appears to be cold gas stripped from the clusters by their motion and pressure effects, and indicates the direction of relative motion.  Region D covers the very hot central region between the clusters.  And Regions C, E, and F were chosen in order to quantify the temperature profile between A115-N and A115-S on either side of the hot region in the center. The spectra were fit with an \textit{APEC} model including photoelectric absorption from galactic neutral hydrogen. The $N_H$ column density was fixed at the value from \citet{starkHI}, 5.2$\times10^{20}$cm$^{-2}$, as in the ACB map fits. The metallicity was left as a free parameter, except in the case of the hot central region, where only poor fits to the metallicity could be obtained. With the exception of the hot central region (region D in Figure \ref{fig:regions}) whose spectral fit we describe later in this section, we fit the spectrum in the energy range 0.7-8.0 keV. The spectral fits are shown in Table \ref{XrayFits}.

In light of the dynamical picture we described above, it is not straightforward to interpret the faint, hot region between the subclusters. Morphologically, this feature is not obviously consistent with either other observed shock features in galaxy cluster X-ray observations, or the appearance of shocks in numerical simulations of galaxy clusters. Interestingly, while working on this manuscript, an X-ray study of Abell 141 appeared that shows morphological similarity to A115 \citep{caglar}. It should be noted that though morphologically this hot region does not appear shock-like, its relative position compared to the two subclusters is where we might expect a shock given the likely dynamical scenario. Since this feature does not appear as prototypically shock-like, we have looked in detail at this feature. We used the ACB temperature map as a guide to discover this, and other features. To verify the temperature in that region, we also have extracted a spectrum from an elliptical region covering the hot region in between the clusters. In the hot region, we modified the spectral fitting slightly, in that we fixed the metallicity at $Z_{\odot}$=0.2, but the temperature fit was relatively insensitive to the choice within the range of the local fitted values around it. Additionally, we fit the spectrum in the 0.7-5.0 keV band, as we see flattening of the spectrum at high energy due to the contribution of the noise. We fit a bulk temperature in this region (marked region "D" in Figure \ref{fig:regions}) of $T_X=11.03 \pm 1.74$~keV.

\begin{table}
\caption{Spectral Temperature Fits}
\vspace{-4mm}
\label{XrayFits}
\begin{center}
\begin{tabular}{|c|c|c|c|}
\hline
Region & Name & T$_X$ (keV) & Z$_{\odot}$ \\
\hline
A &	N Core  &	3.01 $\pm$ 0.03 &	0.29 $\pm$ 0.02 \\
B &	N Tail	& 5.30 $\pm$ 0.10 & 0.22 $\pm$ 0.03 \\
C &	N Inter & 6.90 $\pm$ 0.45& 0.15 $\pm$ 0.06 \\
D & Hot Central & 11.03 $\pm$ 1.7 & fixed \\
E & E Middle & 8.82 $\pm$ 0.78 & 0.46 $\pm$ 0.09 \\
F & S Middle & 7.44 $\pm$ 0.42 & 0.31 $\pm$ 0.07\\
G & S Core & 4.31 $\pm$ 0.15 & 0.29 $\pm$ 0.05 \\
H & S Tail & 4.00 $\pm$ 0.08& 0.28 $\pm$ 0.03\\
\hline\hline
\end{tabular}
\end{center}
\vspace{-6mm}
\begin{center}
\begin{tabular}{l}
Spectral fit for temperature and chemical abundance\\ relative to solar for the regions shown in Figure \ref{fig:regions}. X-ray\\
spectral fitting for these regions is described in Section \ref{sec:StrucAndDyn}.
\end{tabular}
\end{center}
\end{table}
\begin{figure*}[htbp] 
   \centering
   \includegraphics[width=0.7\textwidth]{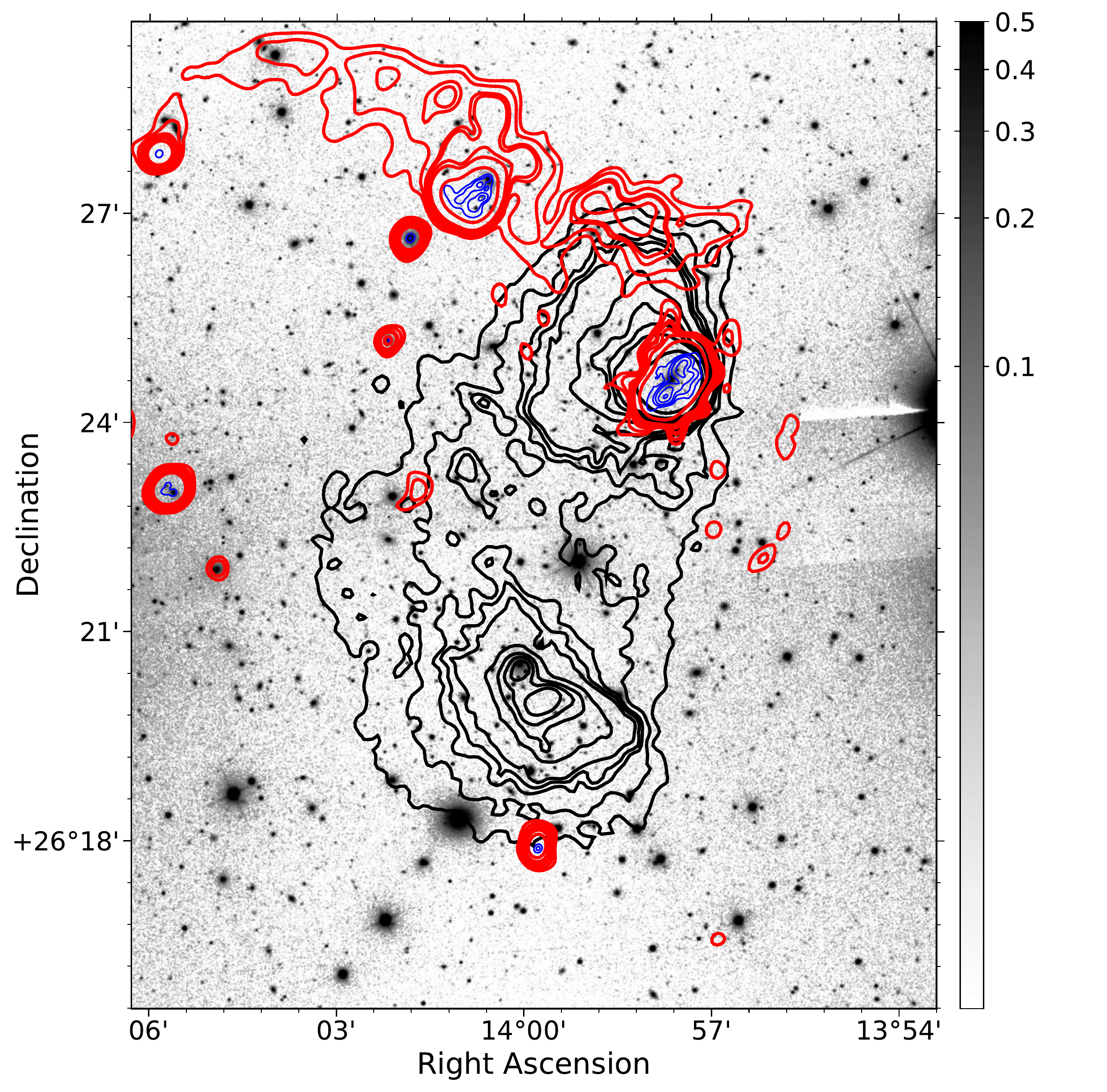}
   \caption{Optical map of A115 using SDSS data, overlaid with 1.5 GHz radio contours showing 3C28, the head-tail source (J0056+2627) and the radio relic. Black contours represent the 0.5-8.0 keV \textit{Chandra} X-ray surface brightness and represent the same values as depicted in figure \ref{fig:acbMap}. Blue contours are constructed with VLA-B data having a beam size of $3.2\arcsec \times 2.8\arcsec$ in position angle 79$^o$ and are at 1.5, 4.25, 12.75, 42, and 85 percent of peak emission. The red contours are from previous VLA C+D array data as published in \citet{Botteon:16} and are at 0.06, 0.10, 0.16, 0.18, 0.26, and 0.40 percent peak emission. The restoring beam of the image is $15\arcsec \times 14\arcsec$ in position angle -35$^o$. The colorbar is in units of nanomaggies (1 nanomaggy = $3.631 \times 10^{ -6}$ Jy).}
   \label{fig:SBandRadio}
\end{figure*}
We have also examined the X-ray emission in the region of the radio relic, as was done in \citet{Botteon:16}. Figure \ref{fig:SBandRadio} shows the X-ray surface brightness and radio contours, overlaid on the optical data. One can clearly see the location of 3C28, and the head-tail radio source (J0056+2627). The red contours, extending from west to east away from the 3C28 at the center of A115-N, are 1.5 GHz  radio contours from the VLA B+D array, showing the location of the extended emission of the radio relic. If the relic is indeed being illuminated because of shock acceleration of a pre-existing population of particles, we need to understand how the presence of a shock at that location is consistent with the dynamics of the cluster. Shock-accelerated particle populations cool quickly by synchrotron radiation, and so the expectation is that the location of the relic will be almost precisely coincident with the shock accelerating the particles.  While we do not see strong evidence for a shock at that location in the X-ray data, it is admittedly very faint. \citet{Botteon:16} make an argument for evidence of a shock from the X-ray data near the radio relic.  However, what is not immediately obvious is what dynamical scenario should result in a shock at that location. We address the dynamics in later sections.

\subsection{Shocks and Temperature Features in the ICM}\label{sec:DynamicsAndShocks}

\begin{figure*}[htbp] 
   \centering
   \includegraphics[width=\textwidth]{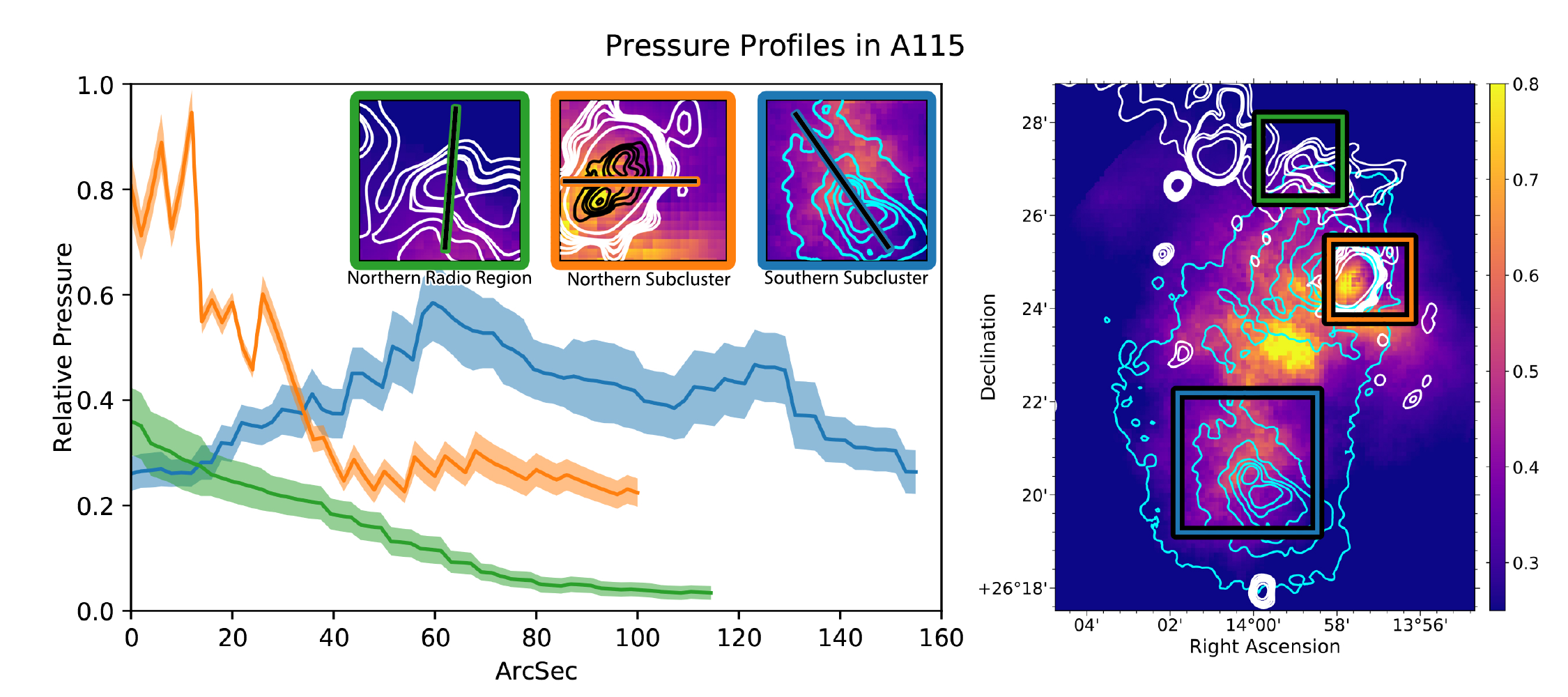}
   \caption{Relative projected pressure profiles from the X-ray surface brightness and temperature maps across 3 interesting features -- upstream of the northern and southern subclusters, and in the location identified by previous studies as a shock across the radio relic feature. Our automated shock finder, described in the text, finds evidence for a shock in the X-ray image in the region to the east of the northern subcluster, where our profile is taken (marked with the orange line). The overview image (right) is of relative pressure difference. X-ray surface brightness contours are depicted in cyan at 1, 1.8, 2.58, 4, and 5.1 percent of peak emission. VLA C+D contour lines are depicted in white at 0.06, 0.10, 0.16, 0.18, 0.26, 0.40 percent of peak emission. VLA B+D contour lines (only shown in the northern subcluster thumbnail) are depicted in black at 1.5, 4.25, 12.75, 42, and 85 percent of peak emission.}
   \label{fig:pressureProfiles}
\end{figure*}

We have run an automated shock finder on the surface brightness and spectral temperature maps, identical to the procedure described in \citet{Datta:14} and \citet{Schenck:14}. High-quality surface brightness and spectral temperature maps can be probed using this technique, which is adapted from a similar calculation used in numerical simulations \citep{ryu03, skillman08}. In numerically simulated clusters, we can find shocks using the full three-dimensional properties of the gas. The conditions for determining whether a given volumetric element of the simulation is the location of a shock are 

\begin{eqnarray}\label{eq:jump_cond}
\nabla \cdot {\bf v} & < & 0 \nonumber \\
\nabla{T}  \cdot \nabla{K_S} & > & 0 \\
 T_2 &>& T_1 \nonumber \\
\rho_2 & > & \rho_1 \nonumber
\end{eqnarray}
where ${\bf v}$ is the velocity field, $T$ is the temperature, $\rho$ is the density and $K_S=T/\rho^{\gamma -1}$ is the entropy. The Mach number of the shock is then defined by the temperature jump, using the Rankine-Hugoniot shock jump conditions, to be

\begin{equation}\label{eq:tempjump}
\frac{T_2}{T_1}=\frac{(5M^2 - 1)(M^2+3)}{16M^2}.
\end{equation}

In X-ray images, all shock observables are projected on the sky. Therefore, for observational data, we must use 2-dimensional projected X-ray surface brightness and temperature maps. Therefore, the technique is modified to account for that. In this modified scenario, the shock-finder calculates the jump in temperature and surface brightness in $N$ evenly placed directions centering on a given pixel. The shock-finder then accepts those pixel-pairs between which the conditions $T_2>T_1$ and $S_{X2} > S_{X1}$ (where $S_{X2}$ and $S_{X1}$ represent the downstream and upstream X-ray surface brightness respectively, and the ratio is a proxy for $\rho_2 > \rho_1$ in observations) in equation~\ref{eq:jump_cond} are satisfied. The other two conditions, $\nabla \cdot {\bf v} < 0$ and $\nabla{T}  \cdot \nabla{K_S} > 0$, in equation \ref{eq:jump_cond} cannot be used in the case of observations. The Mach number for each successful pixel-pair is noted. The Mach number with the maximum value is chosen to be the resultant Mach number for that given pixel.
The full details of the method are described in \citet{Datta:14} and \citet{Schenck:14}.

The automated shock finder identifies only one region where there are pixels consistent with a shock, and that is in the area we interpret as leading the motion of A115-N. The shock finder identifies a number of pixels in the image as shocks, with Mach numbers ranging from $M\approx1$ to $M\approx3$. This area of the map has very low surface brightness, and the temperature map, by the nature of the method, has aggregated pixels over some relatively large area of the map. Again, as before, using the ACB map and shock finder as a guide, below we have explored more deeply with extractions in those regions. 

We explored the region identified as a shock in the above analysis. First, we used a projected pressure map derived from the X-ray surface brightness and temperature maps. The projected pressure maps were generated by taking the square root of the X-ray surface brightness as a proxy for projected density, and multiplied it by the temperature fit to the X-ray spectra, represented by the ACB map.  We then generated a projected pressure profile, across the region identified as a shock, and it is shown in the orange-boxed region and line in Figure \ref{fig:pressureProfiles}. In the region upstream of the northern subcluster, we see a steep drop in pressure.

Other parts of the X-ray map where we might have expected to find evidence of a shock are the region to the northeast of A115-S (leading its motion), and northeast of both subclusters, in the location of the radio relic.   We explored the pressure profile across the northern radio relic in the area where \citet{Botteon:16} suggest they detect a shock.  In Figure \ref{fig:pressureProfiles}, we show the projected pressure profile across that northern radio region (in green).  In front of the southern subcluster (the region shown by the blue box and line), we see no such discontinuity, just a gradual decline in pressure.  This is also true across the northern radio relic region. However, as noted in earlier sections, orientation and projection effects can act to diminish the observed surface brightness and X-ray temperature contrast at the location of a shock.

\begin{figure*}[htbp] 
   \centering
   \includegraphics[width=0.49\textwidth] {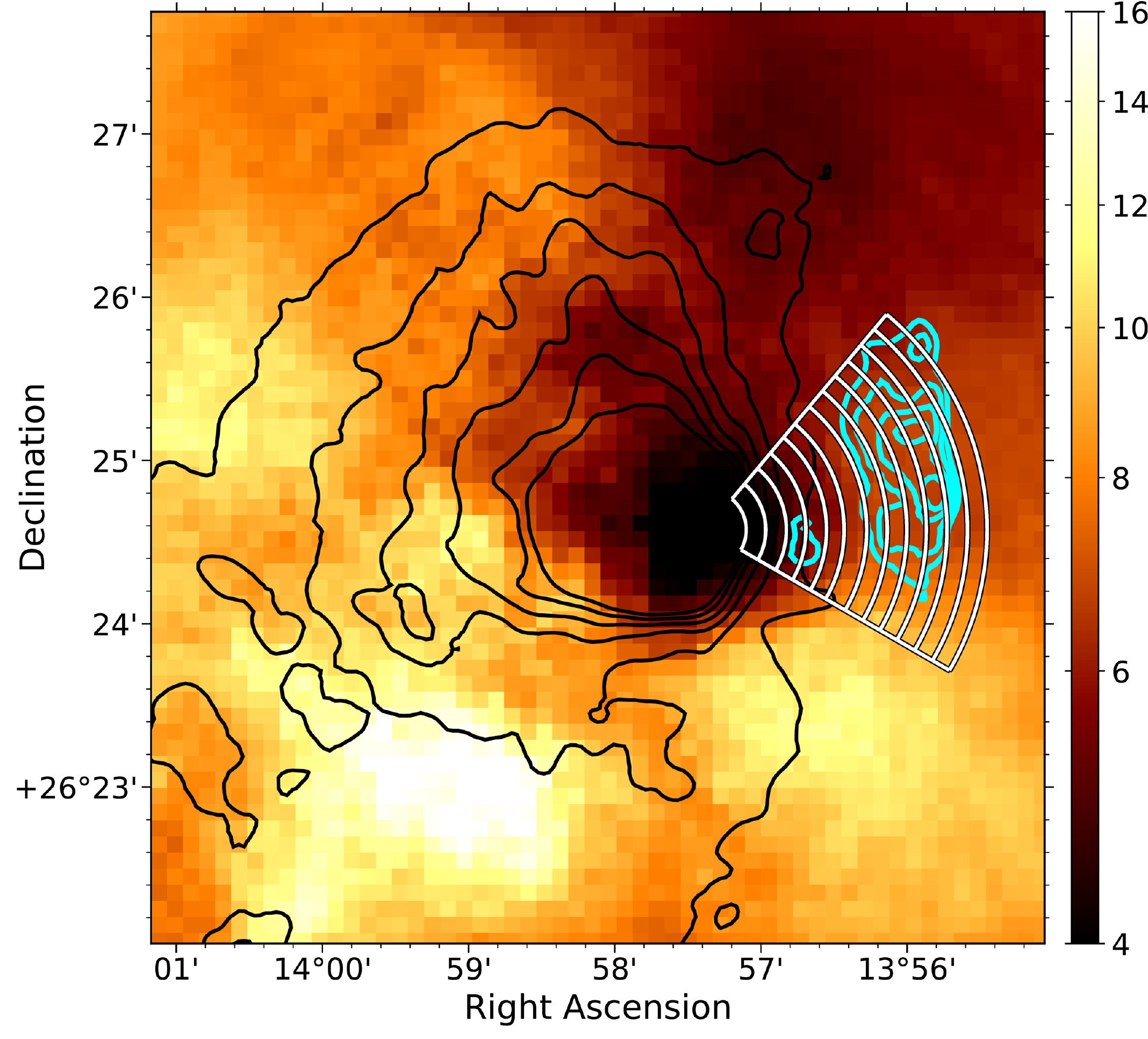}
   \includegraphics[width=0.49\textwidth]{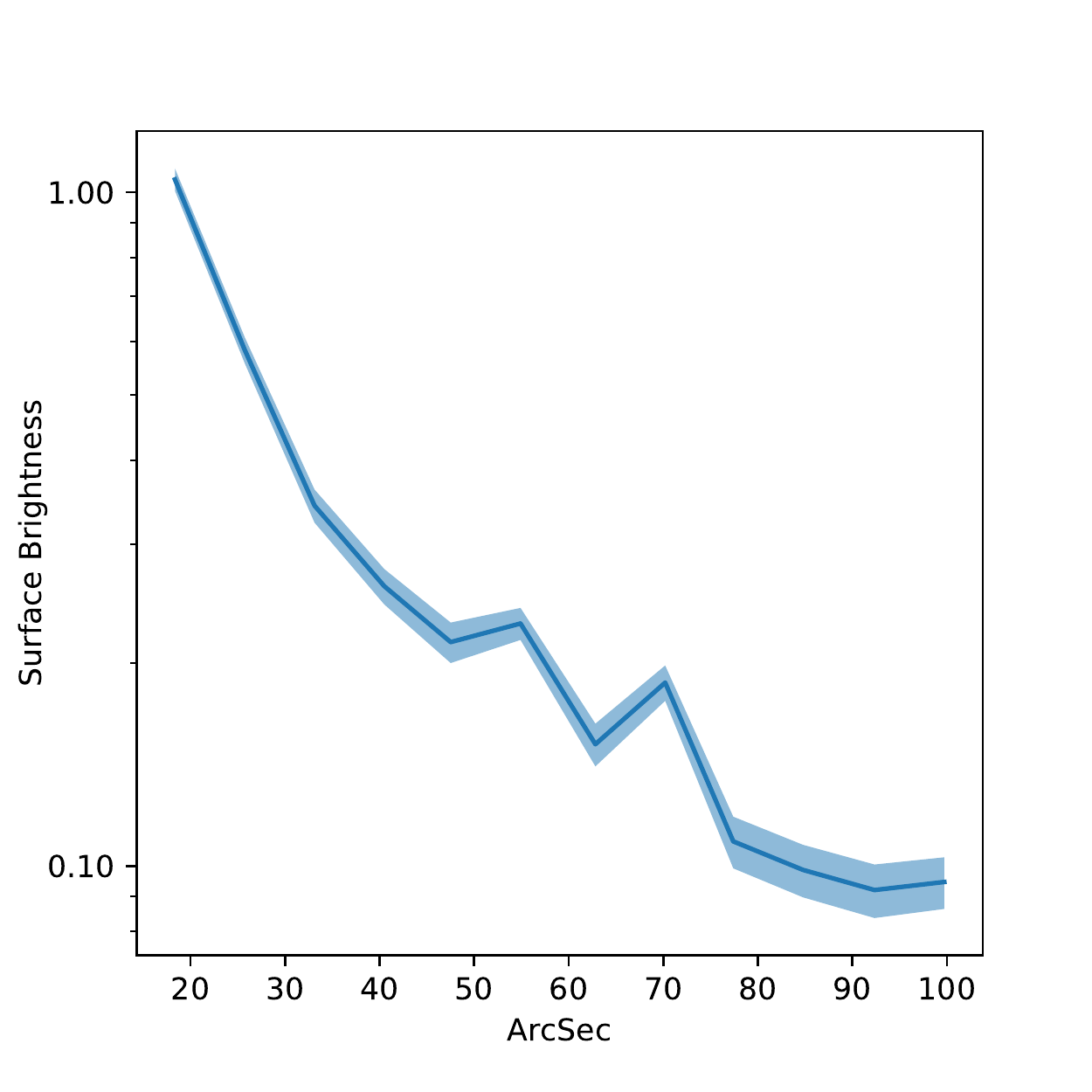}
   \caption{Left panel: ACB Temperature map of the northern subcluster (A115-N) overlaid with contours of X-ray surface brightness (black) and Mach number of identified shocked pixels (cyan). Mach contours are located at 1, 1.25, 1.5, and 1.75. The surface brightness contours are at 1, 1.8, 2.58, 3.38, 4.17, and 4.96 percent of peak emission. White are the annular wedge regions from which the surface brightness is extracted for the radial profile. Right panel: Radial Profile across the feature identified as a shock by the automated shock finder. Note the steep increase in the fifth bin from the right, just behind the identified shock. One $\sigma$ error is shown while the units are in counts per pixel squared.}
   \label{fig:radialProfile}
\end{figure*}

In the direction of motion of A115-N, where the shock detector identifies some pixels as part of a shock, the X-ray surface brightness is quite low.  As a consequence, the region from which pixels are selected for an X-ray spectral fit by both the WVT and ACB methods (both using a signal-to-noise threshold of 50 for their regions) is quite large.  Therefore it is difficult to extract a reasonable spectral fit to the regions either in front of or behind the identified shock location.  However, we can quite easily extract a surface brightness profile across this region ahead of the deduced motion of A115-N, and that is shown in Figure \ref{fig:radialProfile}. The regions from which the surface brightness is extracted are shown in the annular wedge regions in the map also in Figure \ref{fig:radialProfile}.  The fifth region in from the outer part of the shock feature shows a steep increase in surface brightness, by a factor of roughly two. This is the location where the shock finder identifies a shock, and the profile shows that location to be plausible.  Better X-ray data may allow a good spectral fit, spatially resolved in this region.  However, the current data do not permit that at this time.

\begin{figure*}[htbp] 
   \centering
   \includegraphics[width=0.95\textwidth]{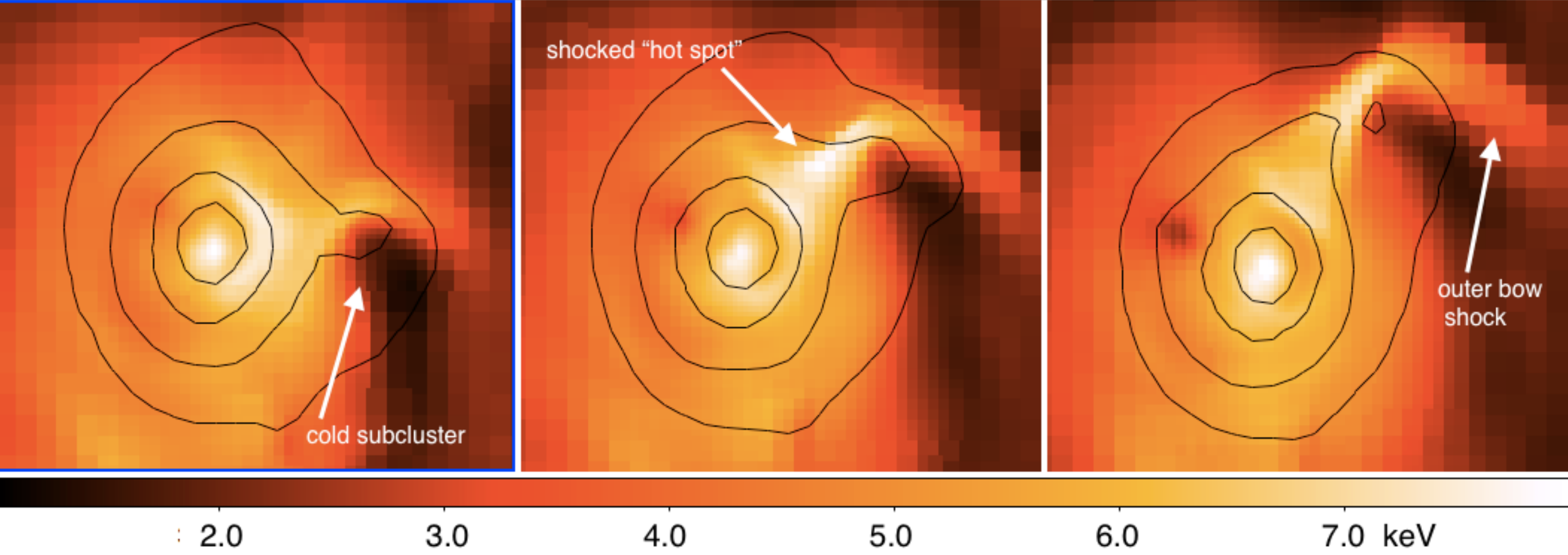}
   \caption{Spectroscopic-like temperature maps of a simulated cluster showing heating between the two subclusters due to bow shock interactions. Simulated 0.3-8.0 keV X-ray contours are overlaid, representing a dynamic range of 1000, similar to the dynamic range of \textit{Chandra} observations. The contours represent (from brightest to faintest) 0.05, 0.03, 0.01 and 0.001 of peak surface brightness.} These three panels are incremented from left to right chronologically. $\delta$t = 220 Myr, and image represents a projected size of about 2$h^{-1}$ Mpc. Note the cold subcluster to the right of each image, moving upward, driving a shock.  The combination of shocks from the relative motion of the main and subcluster results in a hot region between them with relatively flat surface brightness, similar to A115. In the third panel, note the extended bow shock outside the X-ray visible region, similar to what we might be seeing in A115, as described in the text.
   \label{fig:simulatedcluster}
\end{figure*}
\subsection{Comparing to Numerical Simulations}
For comparison, and to help  deduce the dynamics of A115, we show synthetic observations of galaxy clusters simulated in a cosmological context using the cosmological N-body/hydrodynamic simulation code \texttt{Enzo}\footnote{http://enzo-project.org} \citep{enzo}. The merger dynamics in the simulations are an excellent guide for determining the observable consequences of typical subcluster interactions.  The simulations used for this purpose are described in detail in \citet{jeltema}. The simulation used here is of a volume of the Universe 128 $h^{-1}$ Mpc (comoving) on a side, on a $256^{3}$ root grid. The simulation is evolved in a $\Lambda$CDM cosmological model from an \citet{eisenstein} power spectrum from $z=99$ to $z=0$, with a maximum of five levels of adaptive mesh refinement (AMR).  This results in a peak spatial resolution of 15.6$h^{-1}$ comoving kpc. We refine on both dark matter and baryon overdensity of 8.0. This particular simulation does not include the effects of metal line cooling, as has been done in other prior work. In this case, we are using the simulated clusters to understand the larger scale dynamics and observable effects outside the cluster core, where typically cooling times are long. Such simulations serve us as a guide for understanding the range of merger interactions we expect in a cosmological context.

In \citet{jeltema}, we extracted 16 simulated clusters at $z=0$ whose mass exceeded $M_{200} \geq 3 \times 10^{14} M_{\odot}$. The objects of highest virial mass in the simulation are similar in mass to A115. For the current study, we make use of the synthetic observations of these 16 clusters. For each of the identified clusters at $z=0$, we extracted a volume around that cluster in a series of 132 snapshots in time, equally spaced ($\delta t = 0.22$ Gyr) in the redshift interval $0 \leq z \leq 0.9$. For each of these time outputs, using the \texttt{yt}\footnote{http://yt-project.org/} toolkit \citep{yt}, we created synthetic 0.3-8.0 keV X-ray images and spectroscopic-like temperature ($T_{sl}$) maps \citep[see][]{rasia}. The X-ray emission is calculated using the \textit{CLOUDY}\footnote{http://www.nublado.org/} software \citep{ferland}. The projections are of an 8h$^{-3}$ Mpc$^3$ volume centered on the cluster, thus each image is an 8h$^{-2}$ Mpc$^2$ field.  This exposes not only the merger activity within the cluster virial radius, but also the larger cosmological environment of the cluster, allowing us to understand how various observable effects originate. 

\begin{figure*}[htbp] 
\centering
\includegraphics[width=0.7\textwidth]{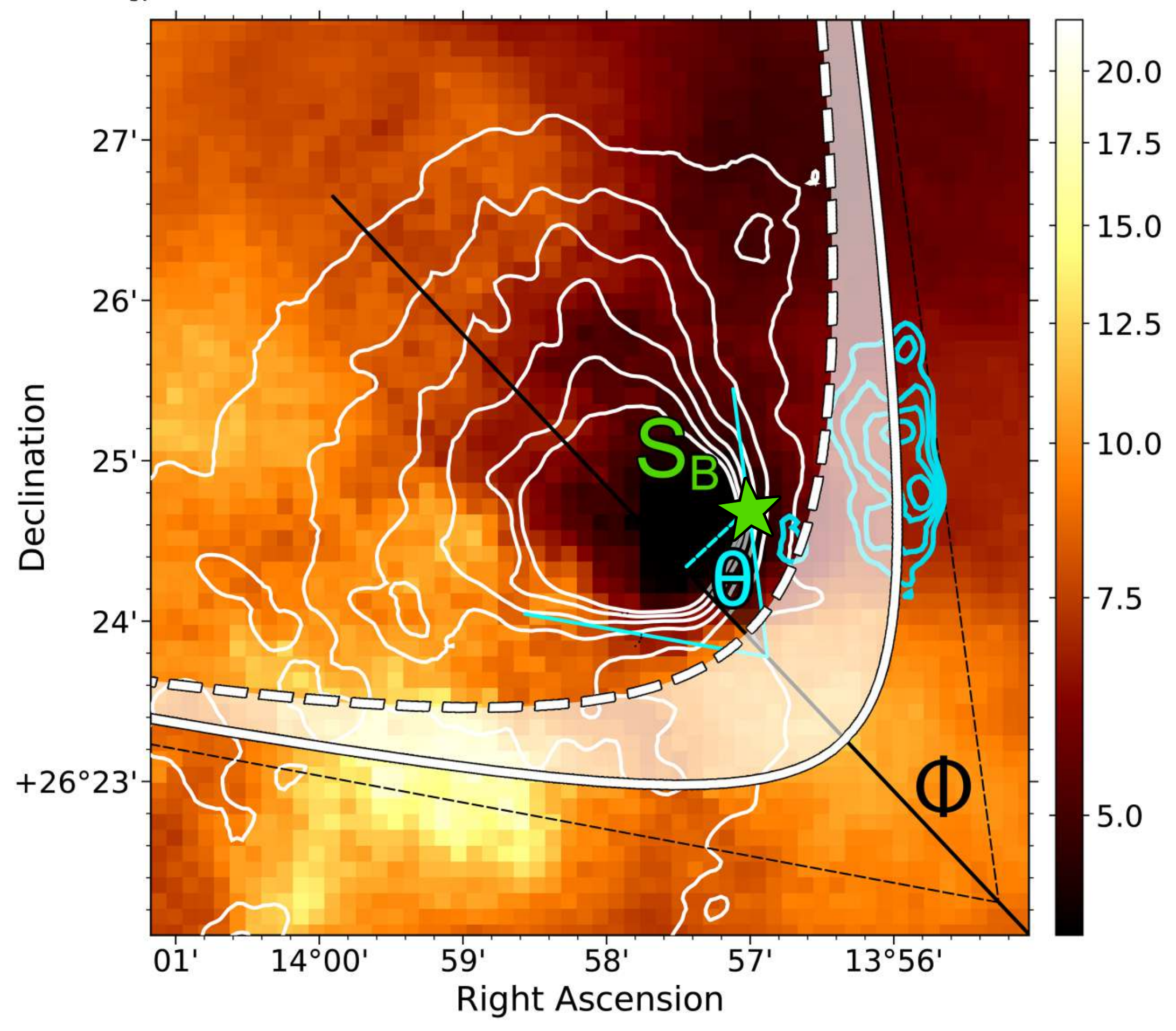}
\includegraphics[width=0.7\textwidth]{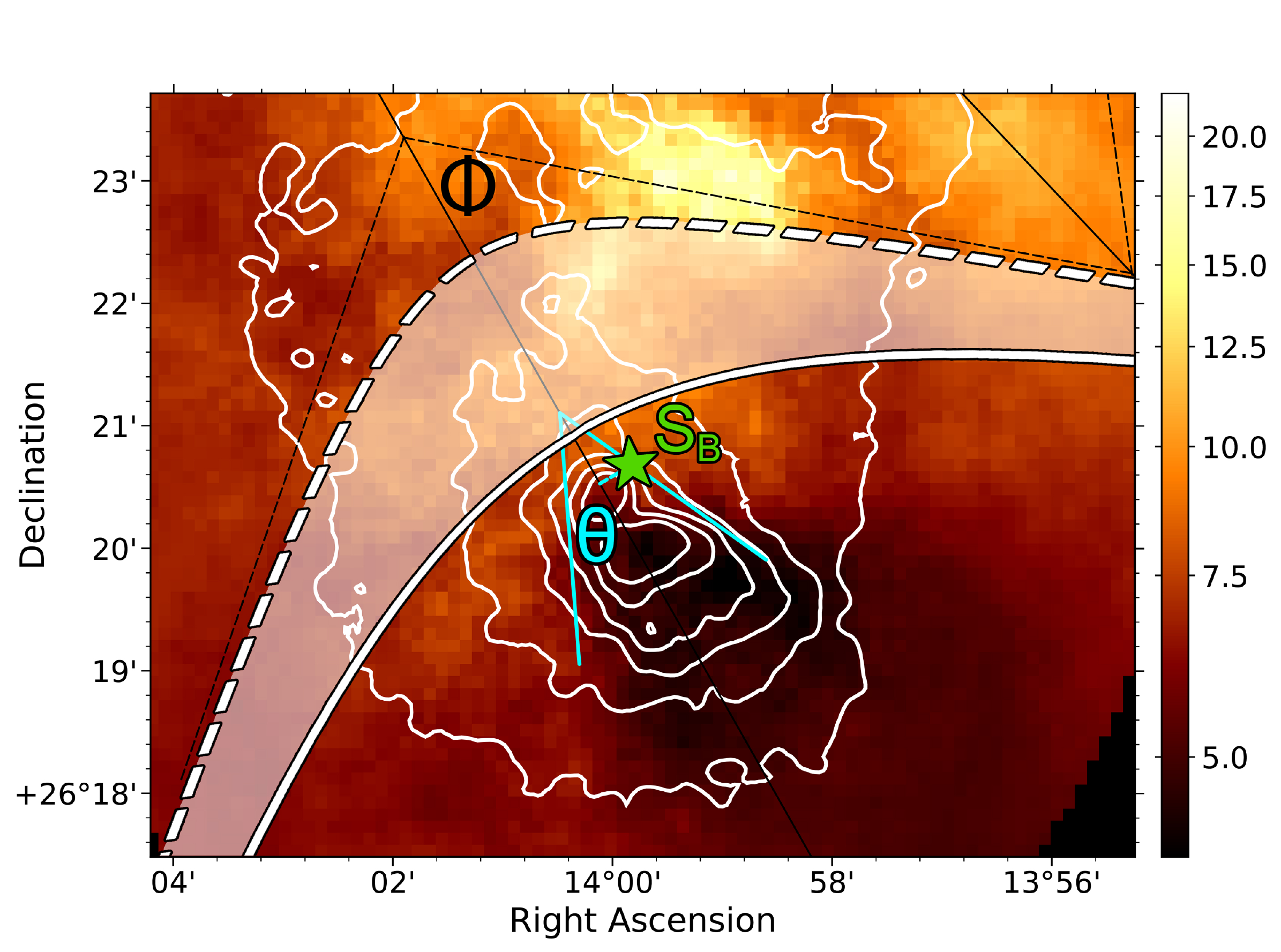}
\caption{Location of the bow shocks in the northern (top panel) and southern (bottom panel) subclusters, as shown by the white shaded regions. Sections~\ref{sec:north} (for the northern subcluster) and~\ref{sec:south} (for the southern; note there the additional assumption taken in this case) describe in detail how we obtain the solid, white hyperbola in each case representing the location of the bow shock when assuming axial symmetry of the subcluster in the direction of movement (solid, black line), through the determination of the angles $\phi$ and $\theta$, and the body sonic point $S_B$ (marked with a green star). The dashed, white hyperbola comes from the same calculation but assuming instead a two-dimensional shape (see M49) of the moving object. We might expect the bow shock to be approximately bracketed by these two curves. These results support the plausibility of our hypothesis that the interaction of the two bow shocks in the center of the cluster drives the observed high temperature that otherwise is unexplained. X-ray surface brightness contours in both images are represented by solid white lines at 1, 1.8, 2.58, 4, and 5.1 percent of peak emission. Both images measured in keV.}
\label{fig:estimateShock}
\end{figure*}

\subsubsection{Similarities to A115 in Numerical Galaxy Clusters}
In the simulation data we see features that are consistent with the inferred dynamics of A115, as well as with the temperature and surface brightness substructure. When two merging subclusters pass each other in the simulations, both drive bow shocks into the other subclusters ICM.  Additionally, these shocks then interact as they travel outward from the subclusters and collide.  This combination of the two subcluster bow shocks can heat the gas between the clusters, leaving an observable structure even after the leading edge of the bow shock has propagated into a less dense, X-ray faint location ahead of the cluster. Figure \ref{fig:simulatedcluster} shows the evolution of such a feature, using the synthetic X-ray temperature maps generated in our prior work. The surface brightness contours represent a dynamic range of 1000 from brightest to faintest, which is similar to the dynamic range of the \textit{Chandra} X-ray images.  The outer edge of the bow shock from the subcluster on the right of the images is in a very faint region of the image, and so would be very difficult to detect. The three panels from left to right are three stages in the cluster evolution, separated by approximately 220 Myr in time. To the right of each image, there is a cooler subcluster traveling toward the top of the image, with a heated region in front of it, representing shock heated gas. Between the subclusters, you can see a hotter region, with a relatively flat surface brightness profile, very similar to what we see in A115. 

Although this is not a model of A115, it may represent a similar dynamical scenario based upon the morphological similarities that we see. These features are quite common in cluster mergers in numerical cosmological simulations, and we believe that a similar scenario is very likely responsible for the hot region between A115-N and A115-S.  It is likely that the interaction of the edge of the two interfering bow shocks has heated the gas to high temperature, and that the upstream parts of the shock, as well as those away from the center of the merger, are in locations where the X-ray surface brightness is too low to detect them.

\subsection{Establishing the Location of the Bow Shock in A115-N}
\label{sec:north}

In order to determine the plausibility of this hypothesis, here we use estimates of the shock Mach number and stand-off distance to see if the location of the central hot region is consistent with this interpretation. We calculate the location of the bow shock around the northern subcluster of A115 using Moeckel's method (M49)\footnote{Technical note available at https://ntrs.nasa.gov/archive/nasa/casi.ntrs.nasa.gov/19930082597.pdf}, which is also summarized for cluster purposes in Appendix B of \citet[][hereafter V01]{Vikhlinin:01}. We employ the shock finder's map of A115 to draw an asymptotic line to the part of the bow shock detected in this manner. This can be seen in the top panel of Figure \ref{fig:estimateShock}. The curved cyan contours in this panel represent the shock contours. The dashed black line represents the asymptotic line of the bow shock. Using the surface brightness map (represented as white contours), we determine the subcluster direction of movement (solid black line). This allows us to obtain the angle $\phi \sim 36$ deg for the northern cluster (for reference, see Fig.~9 in V01), which provides us with a Mach number of the subcluster, $M\sim1.7$, using the relation $\phi=atan(M^2-1)^{-1/2}$.

To calculate the hyperbola function representing the bow shock we require both $M$ and the stand-off distance $x_0$. To obtain the latter, we proceed as follows. Since this object does not have a well-defined shoulder, we can locate the so-called body sonic point $S_B$ (marked with a green star in each panel of Figure~\ref{fig:estimateShock}; and being the point on the surface of the body where the flow speed equals the speed of sound), by using the relation between the angle $\theta$ (formed between the line of movement and the asymptote to the shoulder) and $M$, as found in Fig.~4 of M49. Having this angle, we find $S_B (x_{sb}=549$ kpc$, y_{sb}=103$ kpc$)$ on the surface of the subcluster. The coordinate origin is where the three black lines meet.

From Fig.~7 in M49, selecting the curve that assumes an axially symmetric body with respect to the line of movement, as well as the continuity method, we obtain $L/y_{sb}\sim 1$ (which is only a function of the body speed, i.e. $M$, calculated above to be $\sim1.7$). We can then find the shock detachment distance $L=x_{sb}-x_0$, and finally the shock stand-off distance $x_0 = 445$ kpc.

Given $M$ and $x_0$, we can now model the bow shock as the hyperbola shown by the solid, white line in the {\bf top} panel of Figure~\ref{fig:estimateShock}. Due to the uncertainty in the assumption on the shape of the moving object, we repeat the calculation choosing instead the two-dimensional body, plus continuity method, curve from Fig.~7 of M49. We then obtain $x_0 = 272$ kpc and the dashed, white line in the top panel of Figure~\ref{fig:estimateShock}. We use these two cases to bracket a region (shaded white) where we might expect to find the true, underlying bow shock. \cite{Gutierrez:05} also drew a bow shock estimation from the surface brightness map in their Fig.~7, which is roughly consistent with our result.

Note that we encouragingly find that the bow shock region south of the northern object goes through the hot spot between the two subclusters. We can also obtain the sound speed in the cluster assuming a temperature of 10 keV, $\sim 1594$ km/s, providing $\sim 2710$ km/s for the northern subcluster (using $M \sim 1.7$) (or $1127$ km/s and $1916$ km/s, respectively, assuming 5 keV). Recall that from \cite{Barrena:07} the colliding line of sight velocity between the subclusters is $\sim 1600$ km/s.

\subsection{Estimating the Location of the Assumed Bow Shock in A115-S}
\label{sec:south}

Since for the southern subcluster we do not have any visible residual of the bow shock, we cannot follow the first part of the method used for A115-N. Let us assume though that our argument holds and that the asymptotic line for the southern subcluster is aligned with that of the northern (dashed, black line over the central, hot region between subclusters), as shown in the bottom panel of Figure~\ref{fig:estimateShock}. We also morphologically establish the direction of movement as going through the subcluster as indicated by the solid, black line crossing it. Having these two lines, from this point on we can proceed as described in the previous subsection. For this subcluster, we then obtain $\phi\sim50$ deg and calculate $M\sim 1.3$, which is lower than that of the northern and thus consistent with the dynamical measurements of \cite{Barrena:07}.

To obtain the stand-off distance $x_0$, from Fig.~4 of M49 and using $M\sim 1.3$, we have $\theta\sim 25$ deg. Positioning this angle in the bottom panel of Figure~\ref{fig:estimateShock} (between the solid, black and cyan lines) we obtain $x_{sb}=616$ kpc and $y_{sb}=59$ kpc, and following the same calculation and assumptions as before, $x_0= 528$ kpc. The solid, white line shows the hyperbola corresponding to the assumed bow shock in A115-S, for this being an axially symmetric object. Assuming instead a two-dimensional shape of the subcluster, we find $x_0 = 219$ kpc and the dashed, white line. This and the previous hyperbola bracket the white, shaded region where we might find the bow shock.

\subsection{The Shock Hypothesis}
In summary, the calculations of shock stand-off distance and location strengthen our hypothesis, derived from morphological similarity to numerical simulations, that the location of the central hot spot can plausibly be attributed to the presence of the two bow shocks from the motion of the two subclusters.  Between the subclusters, where there is high enough gas density to generate significant X-ray emission, we can see the hot gas.  In the outer parts of the subclusters, and also upstream of the subcluster motion, the gas density is lower, and the X-ray features we expect for a shock are not visible. 

\section{The Radio Relic and its Relationship to the X-ray}\label{sec:RadioRelic}
Prior work discussing the radio relic in A115 describes two possible scenarios for the origin of the extended radio emission. In \citet{govoni:01} and \citet{Gutierrez:05}, the initial hypothesis was that the extended radio emission is coming from accelerated particles ejected from 3C28 and other nearby radio galaxies, which has subsequently been left behind by the motion of the cluster, and associated galaxies.  The other scenario, described in \citet{Botteon:16}, is that the relic is associated with a shock, accelerating the particles to higher energy, causing them to radiate. These scenarios are not exclusive. This feature could result from a particle population that was ejected from the local radio galaxies and then re-accelerated by a merger shock. Shock acceleration theory (and recent observational evidence) suggests that accelerating a pre-existing population of particles with a nonthermal distribution is more efficient than accelerating particles out of the tail of a thermal distribution \citep{bruJones,reaccelVanWeeren}. 

What is most relevant to this work is determining the plausibility of either scenario, based on the apparent dynamics of the cluster.  Below, we explain the presence of a shock at the location and orientation of the radio relic based on a reasonable dynamical scenario. Also, we investigate whether the radio data support the shock hypothesis, or are consistent with the idea that this is simply stripped radio plasma, with no re-acceleration process. 

\begin{figure*}[htbp] 
   \centering
   \includegraphics[width=0.7\textwidth]{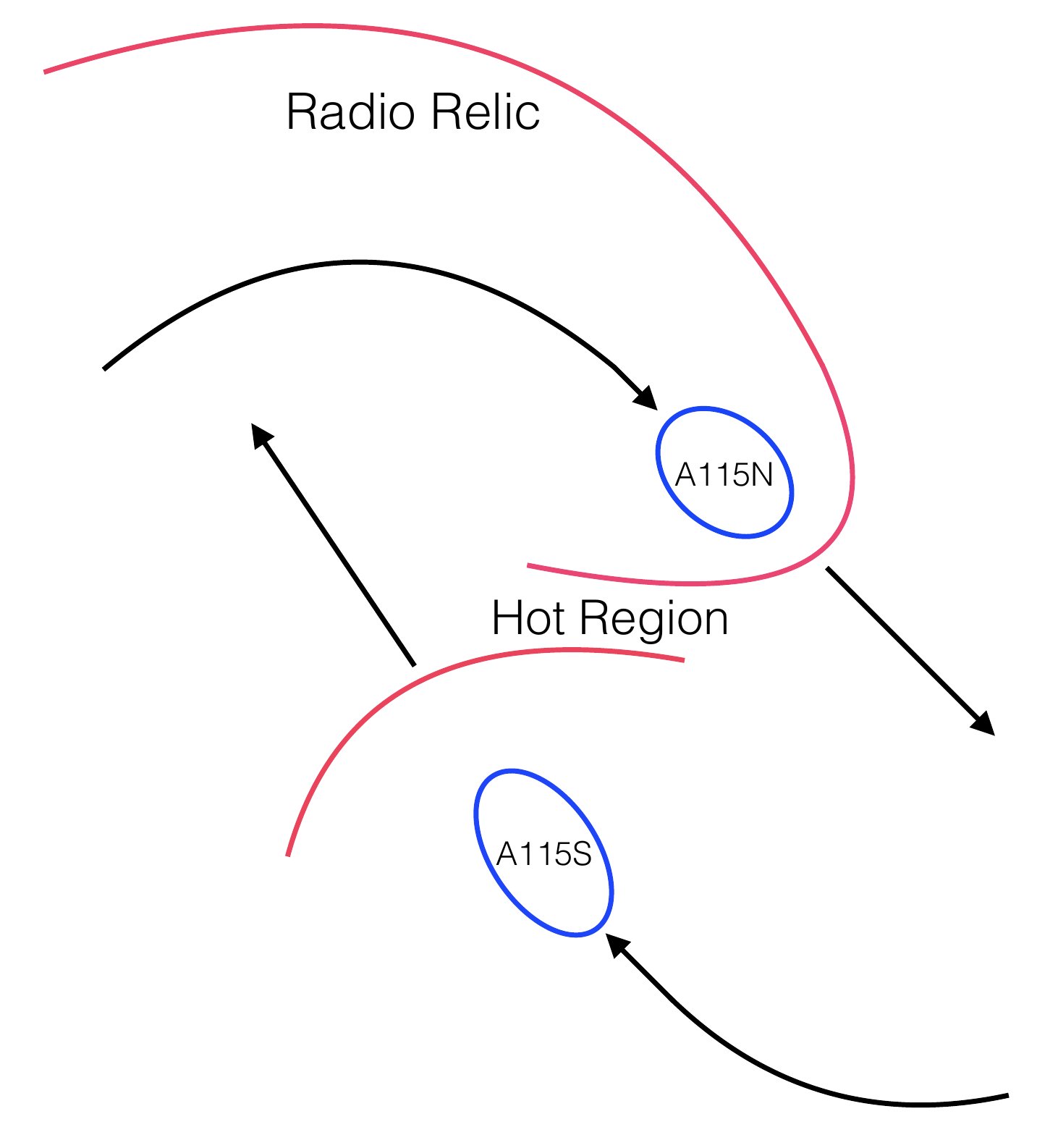}
   \caption{Pictorial representation of the dynamical scenario in A115.  The subclusters are orbiting in each other in the process of merging. Red lines represent the location of shocks, black arrows represent the past and current motion of the subclusters, blue ellipses represent the region of each subcluster core. The radio relic results from the Fermi acceleration of a relic population of cosmic rays ejected from the local radio galaxies, and advected behind the motion of A115N. The curved remnant of the earlier bow shock, similar to what we see in simulations, continues to propagate into the relic plasma after A115N has turned its motion. The interaction of the bow shocks of A115N and A115S produces a hot region between the subclusters.}
   \label{fig:cartoon}
\end{figure*}
The morphology of shocks in merging numerical clusters seems to be in some cases consistent with the shape and location we see in A115. This feature could in fact be consistent with the dynamical picture deduced from X-ray and optical data. In Figure \ref{fig:simulatedcluster}, one can also see the outer part of the bow shock from the cold subcluster.  This region would very likely not be visible in the X-ray observation of such a cluster, as it is in a region of very low X-ray surface brightness. However, should that portion of the bow shock propagate through a region of accelerated particles, one might expect Fermi processes to re-accelerate those particles to higher energy, and they may become visible as synchrotron sources.

As to the second question -- whether the radio data support the shock re-acceleration scenario -- unfortunately the quality of the data prohibit a definitive analysis. What we expect in a shock accelerated scenario is that the radio spectral index will vary spatially, depending on the exact location of the shock. At the location of the shock, the spectral index will be more flat, and behind the shock, as the highest energy particles radiate their energy away more quickly than those at lower energy, the spectral index will steepen.  There are examples from other radio relics, when there is high-quality, high resolution radio data available. If a shock is sweeping through the radio emitting plasma, we would expect a spectral index gradient perpendicular to the long axis of the relic feature \citep[see e.g.,][]{sausage}. However, if this radio emitting plasma is simply stripped from the radio galaxies, and passively advected away, what we would expect is that the spectral index gradient will be parallel to the long axis of the feature, and in the direction of motion of the radio galaxies.

\section{Summary and Conclusions}\label{sec:Discussion}
As discussed in \cite{Barrena:07}, there are a number of features in the optical and X-ray data that support the description of the initial stages of a major merger occurring between the two subclusters of Abell 115. We are able to further corroborate this scenario, consistent with the prior X-ray observations of A115 \citep[e.g.,][]{Forman:81,Shibata:99,Gutierrez:05,Botteon:16}, through more detailed X-ray and radio observations, as well as comparing our results with relevant hydrodynamical simulations.

The dynamical scenario we have described here is shown pictorially in Figure \ref{fig:cartoon}. The subclusters are orbiting in each other in the process of merging. Red lines represent the location of shocks, black arrows represent the past and current motion of the subclusters, blue ellipses represent the region of each subcluster core. The radio relic results from the Fermi acceleration of a relic population of cosmic rays ejected from the local radio galaxies, and advected behind the motion of A115-N. The curved remnant of the earlier bow shock, similar to what we see in simulations, continues to propagate into the relic plasma after A115-N has turned its motion. The interaction of the bow shocks of A115-N and A115-S produces a hot region between the subclusters.

Both prior work, and this analysis, suggest that the A115 merger results in a hot X-ray region between the clusters, which we interpret as a feature resulting from interacting bow shocks.  High-energy particles, accelerated in 3C28 and the head-tail radio source (J0056+2627), are spread out behind A115-N, along the direction of motion.  It has been suggested by \citet{forman17} that hydrodynamic forces associated with the motion of the radio galaxies through the ICM are responsible for the morphology.  Here, we suggest, consistent with the interpretation of the X-ray data by \citet{Botteon:16}, that a shock at the location of the radio relic, in our scenario as part of the leading bow shock of A115-N, has re-accelerated these particles accounting for the radio emission along the arc-like structure between and behind the two radio galaxies.

Though these conclusions are highly plausible, there are some additional observations that could make this case more definitively.  Primarily, the addition of high quality, high resolution, multi-frequency radio observations of the relic could strongly constrain whether the shock acceleration scenario is correct.  In particular, the availability of high-quality, wide bandwidth 350 MHz and 1.4 GHz observations, using the B, C, and D array configurations of the JVLA, would allow for resolved spectral index maps of the relic.  These could be used to determine whether the spectral index gradient is perpendicular to the putative shock, or whether the spectral index variation is more consistent with a scenario where the radio plasma is ejected from the radio galaxies and simply ages via synchrotron cooling as it advects away. 

Also, deep X-ray observations of the region around the relic, as well as in the regions where one might expect the outer parts of the bow shocks resulting in the X-ray hot spot, might clear up that interpretation as well.

\acknowledgments{We thank Andrea Botteon and Fabio Gastaldello for kindly sharing with us the VLA C+D array radio image of A115 which is reproduced here in Figure~\ref{fig:SBandRadio}. This research was supported by NASA ADAP grant NNX15AE17G. We acknowledge NASA's High-End Computing Center for providing computing resources necessary to perform the large number of X-ray spectral fits required for our high-fidelity temperature maps. DR is supported by a NASA Postdoctoral Program Senior Fellowship at NASA's Ames Research Center, administered by the Universities Space Research Association under contract with NASA. \texttt{Enzo} and \texttt{yt} are developed by a large number of independent researchers from numerous institutions around the world. Their commitment to open science has helped make this work possible. AD would like to thank Ramij Raja for help in radio data analysis and imaging.}

\bibliography{references}
\end{document}